\documentclass[aps,prd,reprint,nofootinbib,amsmath,amssymb,superscriptaddress]{revtex4-2} 

\usepackage{graphicx}   
\usepackage[
colorlinks=true,        
citecolor=blue,         
linkcolor=blue,         
urlcolor=blue           
]{hyperref}  
\usepackage{tabulary}
\usepackage{color}      
\usepackage{orcidlink}

\newcommand{\nc}{\newcommand*}  
\newcommand{\beq}{\begin{equation}}
\newcommand{\eeq}{\end{equation}}
\def\({\left(}
\def\){\right)}
\def\[{\left[}
\def\]{\right]}
\nc{\Eq}[1]{Eq.~\eqref{#1}}     
\nc{\Fig}[1]{Fig.~\ref{#1}}     
\nc{\Table}[1]{Table~\ref{#1}}  
\nc{\Sec}[1]{Sec.~\ref{#1}}     
\nc{\red}[1]{\textcolor{red}{#1}}

\begin{document}
\title{Nanohertz Gravitational-Wave Constraints on Supermassive Binary Black Holes at Cosmic Dawn}




\author{Yiqin Chen\orcidlink{0009-0009-0136-7959}}
\thanks{These authors contributed equally to this work.}
\affiliation{Institute for Frontiers in Astronomy and Astrophysics \& Faculty of Arts and Sciences, Beijing Normal University, Zhuhai 519087, China}
\affiliation{School of Physics and Astronomy, Beijing Normal University, Beijing 100875, China}

\author{Shi-Yi Zhao\orcidlink{0009-0001-8885-5059}}
\thanks{These authors contributed equally to this work.}
\affiliation{Institute for Frontiers in Astronomy and Astrophysics \& Faculty of Arts and Sciences, Beijing Normal University, Zhuhai 519087, China}
\affiliation{School of Physics and Astronomy, Beijing Normal University, Beijing 100875, China}

\author{Xingjiang Zhu\orcidlink{0000-0001-7049-6468}}
\email[Contact author: ]{zhuxj@bnu.edu.cn}
\affiliation{Institute for Frontiers in Astronomy and Astrophysics \& Faculty of Arts and Sciences, Beijing Normal University, Zhuhai 519087, China}

\author{N. D. Ramesh Bhat\orcidlink{0000-0002-8383-5059}}
\affiliation{International Centre for Radio Astronomy Research, Curtin University, Bentley, WA 6102, Australia}

\author{Jacob Cardinal Tremblay\orcidlink{0000-0001-9852-6825}}
\affiliation{Max Planck Institute for Gravitational Physics (Albert Einstein Institute), 30167 Hannover, Germany}
\affiliation{Leibniz Universität Hannover, 30167 Hannover, Germany}

\author{Ma\l{}gorzata Cury\l{}o\orcidlink{0000-0002-7031-4828}}
\affiliation{School of Physics and Astronomy, Monash University, Clayton, VIC 3800, Australia}
\affiliation{ARC Centre of Excellence for Gravitational Wave Discovery (OzGrav)}

\author{Valentina Di Marco\orcidlink{0000-0003-3432-0494}}
\affiliation{School of Physics, University of Melbourne, Parkville, VIC 3010, Australia}
\affiliation{ARC Centre of Excellence for Gravitational Wave Discovery (OzGrav)}

\author{George Hobbs\orcidlink{0000-0003-1502-100X}}
\affiliation{Australia Telescope National Facility, CSIRO Space \& Astronomy, PO Box 76, Epping, NSW 1710, Australia}

\author{Wenhua Ling\orcidlink{0009-0009-9142-6608}}
\affiliation{Australia Telescope National Facility, CSIRO Space \& Astronomy, PO Box 76, Epping, NSW 1710, Australia}

\author{Rami F. Mandow\orcidlink{0000-0001-5131-522X}}
\affiliation{Department of Mathematics and Physical Sciences, Macquarie University, NSW 2109, Australia}
\affiliation{Australia Telescope National Facility, CSIRO Space \& Astronomy, PO Box 76, Epping, NSW 1710, Australia}

\author{Richard N. Manchester\orcidlink{0000-0001-9445-5732}}
\affiliation{Australia Telescope National Facility, CSIRO Space \& Astronomy, PO Box 76, Epping, NSW 1710, Australia}

\author{Saurav Mishra\orcidlink{0009-0001-5633-3512}}
\affiliation{Centre for Astrophysics and Supercomputing, Swinburne University of Technology, Hawthorn, VIC 3122, Australia}

\author{Daniel J. Reardon\orcidlink{0000-0002-2035-4688}}
\affiliation{Centre for Astrophysics and Supercomputing, Swinburne University of Technology, Hawthorn, VIC 3122, Australia} 
\affiliation{ARC Centre of Excellence for Gravitational Wave Discovery (OzGrav)}

\author{Sparrow Roch\orcidlink{0000-0002-8713-994X}}
\affiliation{Centre for Astrophysics and Supercomputing, Swinburne University of Technology, Hawthorn, VIC 3122, Australia} 
\affiliation{ARC Centre of Excellence for Gravitational Wave Discovery (OzGrav)}

\author{Christopher J. Russell\orcidlink{0000-0002-1942-7296}}
\affiliation{CSIRO Scientific Computing, PO Box 76, Epping, NSW 1710, Australia}

\author{Ryan M. Shannon\orcidlink{0000-0002-7285-6348}}
\affiliation{Centre for Astrophysics and Supercomputing, Swinburne University of Technology, Hawthorn, VIC 3122, Australia} 
\affiliation{ARC Centre of Excellence for Gravitational Wave Discovery (OzGrav)}

\author{Jingbo Wang\orcidlink{0000-0001-9782-1603}}
\affiliation{Institute of Optoelectronic Technology, Lishui University, Lishui 323000, China}

\author{Shuangqiang Wang\orcidlink{0000-0003-4498-6070}}
\affiliation{Xinjiang Astronomical Observatory, Chinese Academy of Sciences, Urumqi, Xinjiang 830011, China}
\affiliation{Australia Telescope National Facility, CSIRO Space \& Astronomy, PO Box 76, Epping, NSW 1710, Australia}

\author{Andrew Zic\orcidlink{0000-0002-9583-2947}}
\affiliation{Australia Telescope National Facility, CSIRO Space \& Astronomy, PO Box 76, Epping, NSW 1710, Australia}
\affiliation{ARC Centre of Excellence for Gravitational Wave Discovery (OzGrav)}

\collaboration{Parkes Pulsar Timing Array Collaboration}

\begin{abstract}
Standard continuous gravitational-wave searches with pulsar timing arrays (PTAs) neglect cosmological redshift, restricting their applicability to the local Universe.
We introduce a redshift-aware PTA framework and apply it to the Parkes PTA Data Release 3, deriving the first direct constraints on supermassive binary black holes (SMBBHs) at the cosmic dawn.
Evaluating our limits across a broad redshift range, we observationally establish the non-monotonic mass-redshift exclusion boundary driven by the theoretical ``redshift bias", demonstrating how PTAs can effectively probe sources at extreme distances.
Redshift-aware targeted searches toward high-redshift systems, including the ultraluminous quasar J0100+2802 ($z=6.327$) and the JWST-discovered galaxy JADES-GS-z14-0 ($z=14.32$), rigorously exclude SMBBHs with chirp masses $\mathcal{M}_c \gtrsim 10^{10} M_\odot$ across the nanohertz frequency band.
Finally, we demonstrate that high-redshift binaries can be robustly detected and localized, allowing for the accurate measurement of their intrinsic properties. Our results provide the tightest constraints to date on SMBBHs at $z > 6$ and establish a practical framework for probing early-Universe black hole assembly.

\end{abstract}

\maketitle

{\it Introduction.}---Pulsar timing arrays (PTAs)~\citep{Manchester_2013,McLaughlin_2013,Kramer_2013} have opened a new observational window into the population and evolution of supermassive binary black holes (SMBBHs)~\citep{Jaffe_2003,Wyithe_2003,Enoki_2004,Sesana_2008}. Recently, major global PTA collaborations ~\citep{Zic_2023,Agazie_2023_ng15data,Antoniadis_2023,Rana_2025,Chen_2025,Miles_2025} reported strong evidence for a stochastic gravitational-wave background~\citep{Agazie_2023a,Reardon_2023a,EPTA_InPTA_2023,Xu_2023,Miles_2024}, widely thought to be generated by the incoherent superposition of numerous unresolved binaries throughout cosmic history. While this stochastic signal provides a broad demographic census of the overall population, a critical next frontier is the direct observation of continuous gravitational waves (CGWs) from individual exceptionally massive binaries.

Simultaneously, electromagnetic observations have revealed a population of unexpectedly massive early-Universe systems. High-redshift quasars and galaxies at $z \ge 6$ \citep{Harikane_2023,Maiolino_2024,CurtisLake_2023} are now known to host central black holes with masses as high as $10^9 - 10^{10} M_\odot$ \citep{Eilers_2023,Yue_2024,Marshall_2025}, with the most distant spectroscopically confirmed galaxy observed at $z=14.32$ \citep{Carniani_2024}. The existence of these massive black holes at the cosmic dawn raises a fundamental physics question: can current nanohertz GW observatories directly constrain their binary assembly?

Standard CGW searches ~\citep{Agazie_2023b,Antoniadis_2024,Zhao_2025,Zhao_oj287} are inherently restricted to the local Universe because they neglect cosmological redshift, effectively interpreting the observed chirp mass as the intrinsic source-frame mass. However, a decade ago, Rosado et al. \citep{Rosado_2016} theoretically predicted that the detectability of distant SMBBHs by PTAs is governed by a physical effect known as the "redshift bias". Because cosmological redshift reduces the observed frequency while simultaneously mapping to a later, higher-strain stage of the binary’s inspiral, the GW strain amplitude does not decrease monotonically with distance. This counterintuitive physical behavior implies that PTAs could, in principle, constrain exceptionally massive binaries even at extreme cosmological distances.

In this Letter, we provide the first direct observational constraints on high-redshift SMBBHs by realizing this redshift-aware framework using the third data release of the Parkes PTA (PPTA DR3~\citep{Zic_2023}). By translating the global PTA upper limits into source-frame constraints, we observationally establish the non-monotonic mass-redshift exclusion boundary driven by the redshift bias. Applying this formalism to representative targets, such as the ultraluminous quasar J0100+2802 ($z=6.327$) \citep{Wu_2015,Eilers_2023} and the JWST-discovered galaxy JADES-GS-z14-0 ($z=14.32$) \citep{Carniani_2024}, we rigorously exclude equal-mass SMBBHs with intrinsic chirp masses $\mathcal{M}_c \ge 10^{10} M_\odot$. Finally, through signal injection studies, we demonstrate that incorporating a small subset of precise-distance pulsars restores phase coherence in the pulsar term, enabling the robust localization and parameter recovery of high-redshift sources.


{\it Signal model.}---For a circular SMBBH, the observed CGW strain amplitude $h_0$ is given by
\begin{equation}
h_0 = \frac{2 (\pi f_\mathrm{GW})^{2/3} (G\mathcal{M}_z)^{5/3}}{c^4D_L},
\label{eq:strain}
\end{equation}
where $f_\mathrm{GW}$ is the observed GW frequency, $\mathcal{M}_z$ is the redshifted chirp mass, and $D_L$ is the luminosity distance.
The redshifted chirp mass and observed GW frequency are related to their source-frame counterparts through $\mathcal{M}_z = (1+z)\mathcal{M}_c$ and $f_\mathrm{GW} = f_\mathrm{src}/(1+z)$, respectively.
For a nearly monochromatic PTA signal, these relations imply that redshift cannot in general be determined independently from the GW data alone, since the observables constrain redshifted rather than source-frame quantities. Here, the intrinsic (source-frame) chirp mass
\begin{equation}
\mathcal{M}_c=\left[\frac{q}{(1+q)^{2}}\right]^{3/5}M_\mathrm{tot},
\label{eq:chirp mass}
\end{equation}
for a binary with mass ratio $q$ and total mass $M_{\rm tot}$, governs the orbital evolution of the system. 

Conventional CGW searches effectively treat the inferred chirp mass directly as an intrinsic source property. While this approximation is valid in the local Universe, it systematically misidentifies the mass scale of binaries at cosmic dawn. To rigorously constrain early-Universe systems, our framework maintains explicit redshift dependence. In the absence of a detection, a targeted search yields a 95\% upper limit on the GW strain amplitude, $h_0^{\rm UL}(f)$. We directly translate this into a constraint on the source-frame chirp mass:
\begin{equation}
\mathcal{M}_c^{\rm UL}(f,z)=\frac{1}{1+z} \left[ \frac{h_0^{\rm UL}(f)\,c^4\,D_L} {2(\pi f)^{2/3}G^{5/3}} \right]^{3/5}.
\label{eq:McUL}
\end{equation}

Equation~(\ref{eq:McUL}) captures the fundamental interplay between cosmic expansion and binary orbital dynamics. While the luminosity distance $D_L$ geometrically suppresses the observed strain, cosmological redshift dictates that for a fixed observed frequency, a more distant source is emitting at a higher rest-frame frequency $f_{\rm src} = f_{\rm GW}(1+z)$. Because a binary emits stronger GWs during these later stages of inspiral, high-redshift sources are intrinsically louder. At $z \gtrsim 2.6$, this intrinsic amplitude enhancement overtakes the $1/D_L$ distance suppression. Consequently, the exclusion boundary $\mathcal{M}_c^{\rm UL}(f,z)$ is non-monotonic, turning over and allowing PTAs to efficiently probe the high-redshift Universe. This physical mechanism is the core of the ``redshift bias" \citep{Rosado_2016}. To demonstrate this effect across the epoch of early black-hole assembly and recent JWST discoveries, we map this constraint surface over the redshift range $0.01 \le z \le 20$.

{\it Binary black hole mass constraints at cosmic dawn.}---We analyze the PPTA DR3 dataset, comprising high-precision timing observations of 32 millisecond pulsars spanning approximately 18 years. To ensure consistency with recent GW analyses, we exclude PSRs J1824-2452A and J1741+1351, both of which provide negligible sensitivity to GWs. Pulse times of arrival are referenced to the TT(BIPM2020) timescale and processed using the DE440 Solar System ephemeris. The timing noise is modeled using the full PPTA DR3 noise prescription, including pulsar-specific white and intrinsic red noise, alongside a common red-noise (CRN) process~\citep{Reardon_2023b}.

We perform targeted searches toward two extreme early-Universe systems motivated by electromagnetic observations. The first is the ultraluminous quasar J0100+2802 ($z = 6.327$), which hosts a central black hole with an estimated mass of $\sim 1.2 \times 10^{10} M_\odot$~\citep{Wu_2015,Eilers_2023}. The second is JADES-GS-z14-0 ($z = 14.32$), currently the most distant spectroscopically confirmed galaxy, with an inferred stellar mass of $\log_{10}(M_{\mathrm{star}}/M_\odot)=8.6^{+0.7}_{-0.2}$ from spectral energy distribution modelling~\citep{Carniani_2024}. 


For both targets, we carry out targeted searches by fixing the source sky position to the electromagnetic coordinates reported in the literature~\citep{Eilers_2023,Carniani_2024}. We also use the observed spectroscopic redshift of each target to infer its luminosity distance under the adopted cosmology~\citep{Planck_2018}, and keep this distance fixed in the search. By leveraging external electromagnetic data to fix the source sky location, luminosity distance and redshift, our search is expected to improve upon an all-sky analysis.
For the remaining parameters, we adopt the same priors as in the PPTA DR3 all-sky CGW search~\citep{Zhao_2025}.

\begin{figure}[htbp]
    \begin{center}
        \includegraphics[width=1.0\columnwidth]{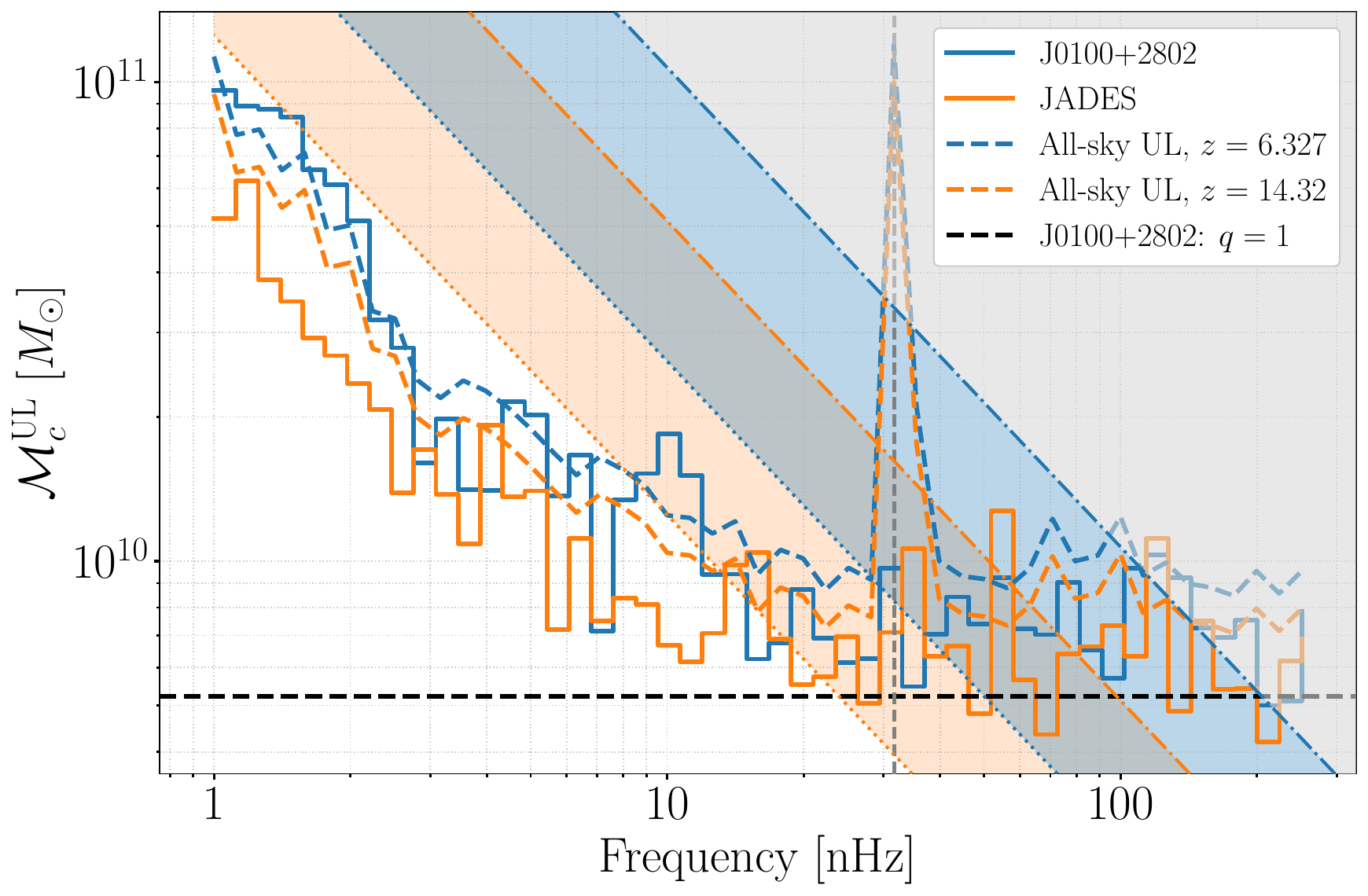}
        \caption{Frequency-dependent 95\% upper limits on the intrinsic chirp mass $(\mathcal{M}_c^{\rm UL})$ derived from targeted searches toward the ultraluminous quasar J0100+2802 at $z=6.327$ (blue solid curve) and the JWST-discovered galaxy JADES-GS-z14-0 at $z=14.32$ (orange solid curve). 
        For comparison, the dashed curves show the PPTA DR3 all-sky strain upper limits converted into source-frame chirp-mass limits using Eq.~(\ref{eq:McUL}) with the redshift and luminosity distance of each target. No target sky position is used in this conversion; the sky positions enter only the targeted searches shown by the solid curves.
        The black dashed horizontal line denotes the source-frame chirp mass $(\mathcal{M}_c \simeq 5.2 \times 10^9 M_\odot)$ for an equal-mass binary ($q=1$) inferred from the reported $1.2 \times 10^{10} M_\odot$ central black hole mass of J0100+2802. Note that the sharp sensitivity degradation at $f = 1/\text{yr}$ ($\simeq 31.7$ nHz) in the all-sky limits arises from covariance with the pulsars' astrometric timing-model fits; conversely, our targeted searches retain full sensitivity across this frequency because fixing the source coordinates breaks this global geometric degeneracy.
        The blue and orange shaded bands denote the frequency intervals bounded by the ISCO and ringdown scales for J0100+2802 and JADES-GS-z14-0, respectively.
        The dotted and dash-dotted boundaries mark the observed ISCO and ringdown frequencies.
        Parameter space residing at frequencies beyond the ringdown scales corresponds to systems that have already merged and no longer emit GWs (see Supplemental Material~\cite{supplemental_full}).
       } 
        \label{mass_ul}  
    \end{center}
\end{figure}

Applying our redshift-aware framework to the PPTA DR3 dataset, we derive frequency-dependent 95\% upper limits on the source-frame chirp mass for J0100+2802 and JADES-GS-z14-0 (\Fig{mass_ul}). Driven by the array's strain sensitivity, the constraints tighten significantly at higher frequencies, flattening to $\mathcal{M}_c^{\rm UL} \sim \text{few} \times 10^9 M_\odot$ in the most sensitive band. Notably, despite JADES-GS-z14-0 residing at a significantly higher redshift, its mass limits are comparable to, and at some frequencies tighter than, those of J0100+2802. This non-trivial scaling perfectly illustrates the redshift bias in action.

Crucially, for J0100+2802, our limits approach the reference source-frame chirp mass for an equal-mass binary ($\mathcal{M}_c \simeq 5.2 \times 10^9 M_\odot$) inferred from its reported central black-hole mass. This confirms that current PTA datasets have already crossed the threshold necessary to physically constrain the binary fraction of the most massive early-Universe systems.
Conversely, given the relatively low inferred stellar mass of JADES-GS-z14-0 ($M_{\rm star} \sim 10^{8.6} M_\odot$), its central black hole mass is expected to be orders of magnitude below our current upper limit. Nevertheless, this target serves as a powerful proof-of-concept, demonstrating that redshift-aware PTA constraints can already be formulated for systems deep into the cosmic dawn.

For high-redshift binaries, our CGW constraints at the highest frequencies are physically bounded by the innermost stable circular orbit (ISCO) and subsequent ringdown scales. 

We explicitly mark these mass-dependent boundaries in \Fig{mass_ul}. Since both the ISCO and ringdown frequencies scale approximately as $[M_{\rm tot}(1+z)]^{-1}$, they appear as slanted transition bands in the $(f_{\rm GW},\mathcal{M}_c)$ plane. The region between the ISCO and ringdown boundaries corresponds to the plunge-merger-ringdown transition, where the standard long-lived inspiral CGW assumption is no longer directly applicable.

\begin{figure}[t]
    \begin{center}
        \includegraphics[width=1.0\linewidth]{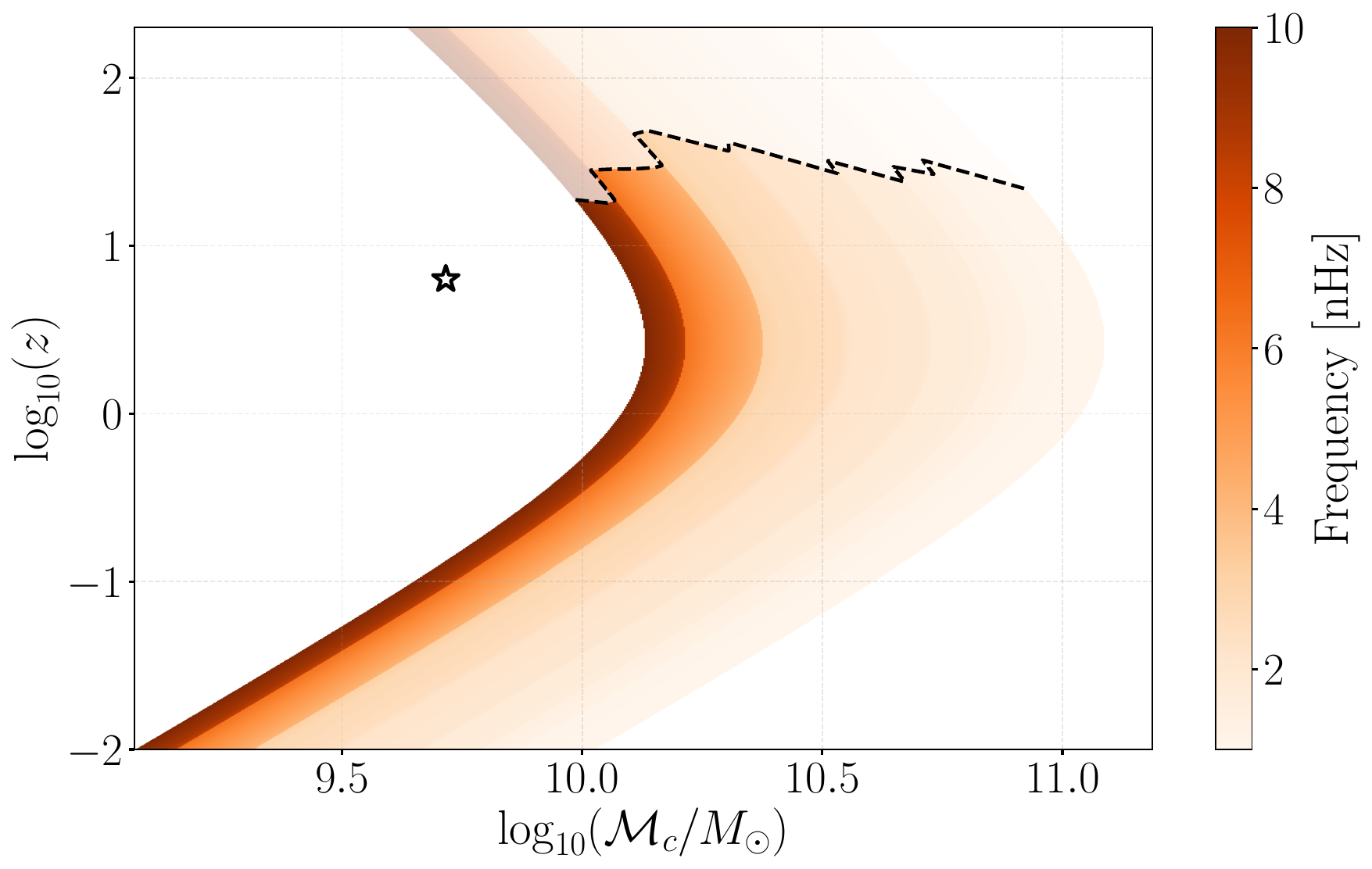}
        \includegraphics[width=1.0\linewidth]{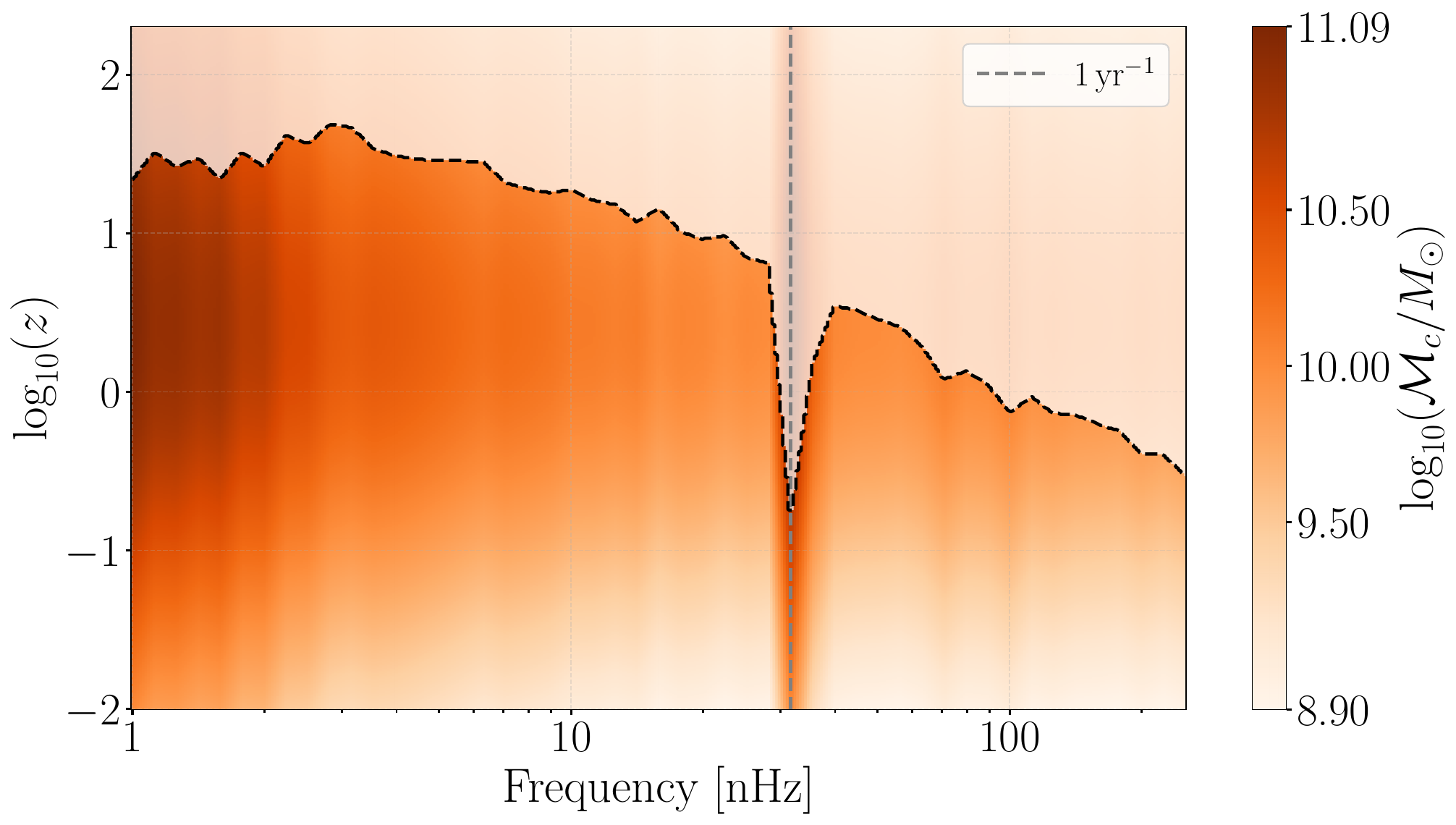}
        \caption{Upper limits on the intrinsic chirp mass of SMBBHs from PPTA DR3, shown in two complementary projections.
        \textbf{Upper panel:} The constraint on the binary chirp mass as a function of redshift over the $1$--$10$ nHz range, with color denoting the observed GW frequency $f_{\rm GW}$. Constant-color bands therefore correspond to fixed-frequency slices of the lower panel. The star marks the J0100+2802 equal-mass reference point, $z=6.327$ and $\mathcal{M}_c=5.2\times10^9\,M_\odot$, as in \Fig{mass_ul}. The black dashed curve marks the ISCO boundary; the shaded region beyond it is outside the physical applicability of the long-lived inspiral CGW model. Its jagged appearance reflects the discrete, frequency-dependent PPTA DR3 upper limits.
        \textbf{Lower panel:} The constraint on the chirp mass shown in the $\left[f_{\rm GW},\log_{10}(z)\right]$ plane, with color denoting $\log_{10}(\mathcal{M}_c^{\rm UL}/M_\odot)$. At each point, binaries with $\mathcal{M}_c>\mathcal{M}_c^{\rm UL}(f_{\rm GW},z)$ are excluded, while smaller chirp masses are not. Smaller $\mathcal{M}_c^{\rm UL}$ values indicate stronger constraints. The vertical dashed line marks $f = 1/\text{yr}$ ($\simeq 31.7$ nHz), where the all-sky strain sensitivity is degraded by covariance with annual astrometric timing-model terms.
        Together, the two panels illustrate how PPTA DR3 constrains massive SMBBHs over the nanohertz frequency band across a wide redshift range.}
        \label{global_UL}  
    \end{center}
\end{figure}

The two panels of \Fig{global_UL} present the same set of upper-limit results from two complementary ways. The lower panel shows the 95\% upper limit on the intrinsic chirp mass, $\log_{10}(\mathcal{M}_c^{\rm UL}/M_\odot)$, as a function of observed GW frequency $f_{\rm GW}$ and redshift z, with values encoded by color. The upper panel focuses on the $1$–$10$ nHz sensitivity band and instead displays the chirp mass limit as a function of redshift $z$. The dashed ISCO boundary marks the limit of physical applicability for the long-lived inspiral continuous-wave model, and should not be interpreted as an additional observational exclusion contour.

As demonstrated in \Fig{mass_ul}, our targeted constraints are only moderately tighter than the corresponding all-sky bounds. Consequently, re-scaling the all-sky limits provides a conservative approximation for mapping the array's sensitivity across a broad range of distances. To place these limits into a global cosmological context, we translate the PPTA DR3 all-sky upper bounds into a constraint surface over the redshift range $0.01 \le z \le 20$ (\Fig{global_UL}).
This demonstrates the non-monotonic exclusion boundary in the mass-redshift plane. 
At sufficiently high redshift, the intrinsic amplitude enhancement of the source overtakes the suppression from the luminosity distance. As a result, the chirp mass required for a detectable signal turns over, allowing PTAs to retain strong sensitivity to extreme-mass binaries well into the epoch of cosmic dawn.
We explicitly truncate these limits at the ISCO boundary; binaries residing above this curve in the mass-redshift plane have already plunged and merged, rendering them physically absent from the continuous-wave inspiral band.

{\it Phase-coherent source localization of high-redshift binaries.}---
After deriving upper limits for high-redshift SMBBHs from both targeted and all-sky PPTA DR3 searches, 
we now turn to the complementary detection problem: if a J0100+2802-like SMBBH signal were present in future PTA data, how accurately could its intrinsic parameters and sky position be recovered?
To evaluate the detectability of early-Universe SMBBHs and the physical role of pulsar-term phase coherence, we perform a targeted signal injection study.
We construct a 40-year mock dataset comprising 32 pulsars, including 30 pulsars from the PPTA DR3 and two additional pulsars, J0636$-$3044 and J2222$-$0137, from the MeerKAT 4.5-year dataset~\citep{Miles_2025}.
The observing cadence for each pulsar is extrapolated from its actual times of arrival in the corresponding dataset, and the noise realizations are generated using the specific white- and red-noise parameters derived for each pulsar.
Into the realistic noise background, we inject a CGW signal corresponding to an equal-mass binary at the precise sky location of J0100+2802 ($z=6.327$, $\mathcal{M}_c \simeq 5.2 \times 10^9 M_\odot$, $f_{\rm GW} = 10^{-8.5}$ Hz), yielding a network signal-to-noise ratio of 23.25.
For signal parameter estimation, the luminosity distance $D_L$ is inferred from the GW signal, and the corresponding redshift is derived from $D_L$ using the distance--redshift relation of the adopted cosmological model.
Further details of the injection study can be found in Supplemental Material~\cite{supplemental_full}.

We analyze the mock dataset using two distinct Bayesian search strategies. The first is a standard search, which treats pulsar distances as independent nuisance variables, leaving source localization largely reliant on the array's broad antenna response pattern~\citep{Taylor_2026}. The second is a ``phase-linked" search that incorporates precise distance measurements for a small subset of six pulsars (J0030+0451, J0437$-$4715, J1744$-$1134, J1909$-$3744, J0636$-$3044, and J2222$-$0137), following the anchor-pulsar strategy of Wen et al.~\citep{Wen_2026}, to restore Earth-pulsar-term phase coherence.
This strategy is similar to the CGW localization studies of Kato and Takahashi~\citep{Kato_2023,Kato_2026}, which showed that accurate pulsar distances can constrain pulsar-term phases and substantially improve continuous-wave source localization. Here we apply the ``anchor-pulsar" localization scheme to a redshift-aware high-redshift SMBBH signal at the position and redshift of J0100+2802.

\begin{figure*}[t]
    \begin{center}
        \includegraphics[width=0.9\linewidth]{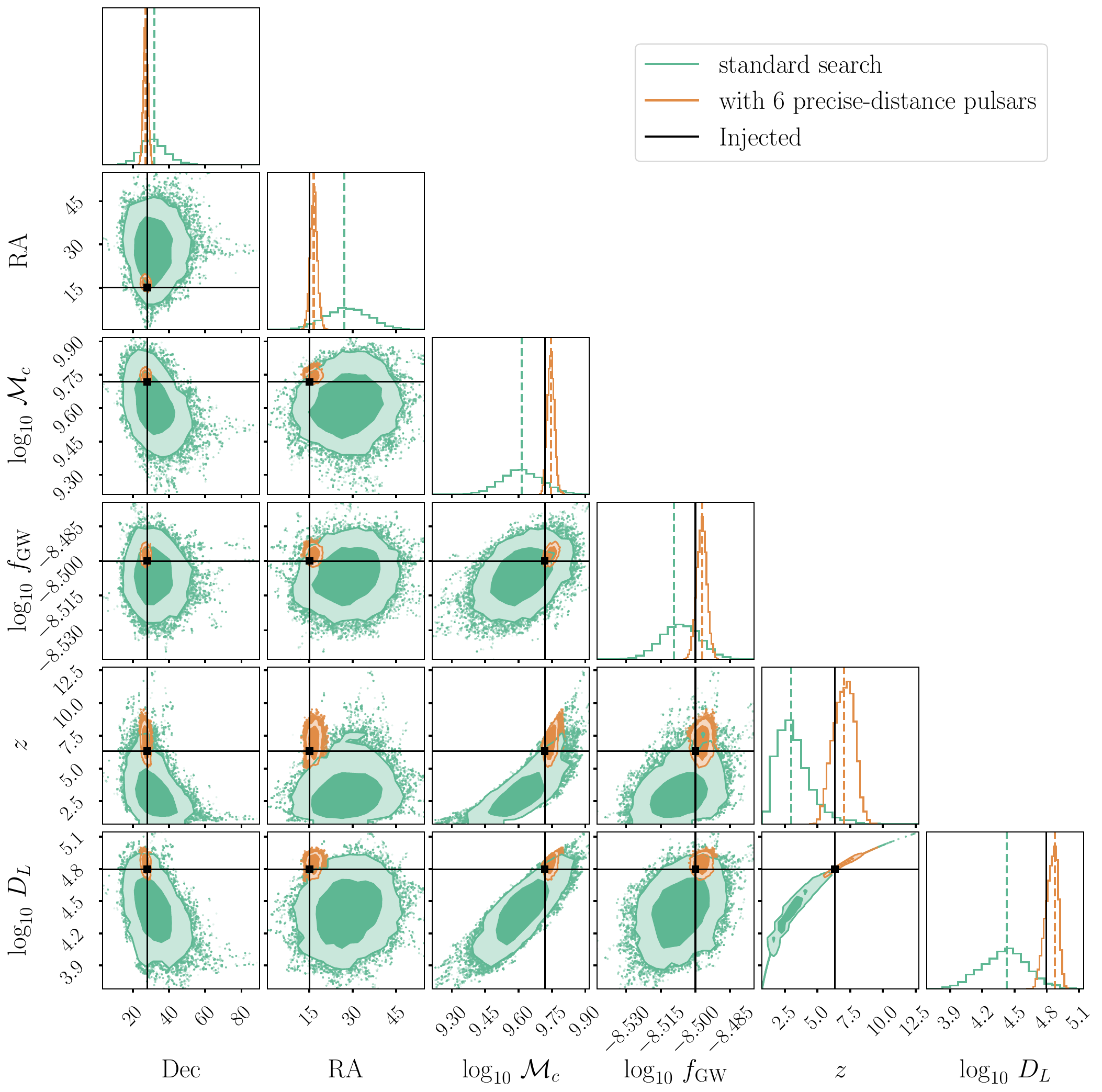}
        \caption{Marginalized one- and two-dimensional posterior distributions for a simulated J0100+2802-like SMBBH injected into a 40-year 32-pulsar mock dataset. The injected CGW signal (black solid lines and markers) corresponds to $\log_{10}(\mathcal{M}_c/M_\odot) \simeq 9.7$, $\log_{10}(f_{\rm GW}/{\rm Hz}) = -8.5$, and $z=6.327$ at the sky position of J0100+2802. Green histograms and contours denote the results of a standard phase-decoupled search, where the source sky localization relies on the array's broad antenna response pattern. Orange histograms and contours illustrate a phase-linked search incorporating precise distance priors for a six-pulsar subset (J0030+0451, J0437$-$4715, J1744$-$1134, J1909$-$3744, J0636$-$3044, and J2222$-$0137). 
        The displayed redshift posterior is derived from the luminosity-distance posterior using the distance--redshift relation under the adopted cosmology~\citep{Planck_2018}. Note that the distance and redshift parameters are strongly degenerate; this effect is most pronounced in the standard search, whose broad posterior extends toward lower redshifts, placing the injected values near the edge of the $2\sigma$ credible region. The inner and outer solid contours correspond to the $1\sigma$ and $2\sigma$ credible levels, respectively. In the joint posterior of the source-frame chirp mass and redshift for the phase-linked search, the injected value lies just outside the $2\sigma$ contour, which is an expected statistical consequence of the specific noise realization. Dashed vertical lines indicate the maximum-a-posteriori estimates.}
        
        \label{corner_compare}  
    \end{center}
\end{figure*}

As shown in \Fig{corner_compare}, explicitly locking the relative phase of the passing wavefront via rapid angular interference oscillations not only sharpens the posterior constraints on the intrinsic chirp mass and luminosity distance, but fundamentally transforms the array's localization capability.
Quantitatively, the 90\% credible sky area collapses from $567.78~{\rm deg}^2$ to $19.09~{\rm deg}^2$, corresponding to a factor of over 30 reduction. Such a dramatic collapse demonstrates that phase-linking transitions continuous-wave localization from broad hemispheric constraints to targeted search fields, providing the high-precision astrometry critical for future multi-messenger campaigns.

{\it Conclusion.}---In this Letter, we have established the first direct observational bridge between nanohertz gravitational-wave astronomy and the cosmic-dawn Universe revealed by JWST. By explicitly incorporating cosmological redshift into the continuous-wave signal model, we overcome the local-Universe limitations of standard PTA searches to derive the first direct constraints on the intrinsic chirp masses of early-Universe SMBBHs using the PPTA DR3.

Applying this redshift-aware framework, we rigorously excluded equal-mass binaries with $\mathcal{M}_c \gtrsim 10^{10} M_\odot$ in extreme targets such as J0100+2802 ($z=6.327$) and JADES-GS-z14-0 ($z=14.32$). Crucially, by mapping the all-sky constraint surface, we provided the first observational realization of the theoretical redshift bias predicted by Rosado et al. \citep{Rosado_2016}. This confirms that the intrinsic strain enhancement of late-stage inspirals at high redshift can effectively overcome cosmological distance suppression, granting PTAs a non-monotonic sensitivity window into the early Universe.

Furthermore, our injection studies delineate a clear methodological path toward the unambiguous detection and localization of these distant sources. We demonstrated that treating pulsar-term phases as decoupled nuisance variables leaves standard searches limited by the array's broad antenna response pattern. Conversely, adopting a phase-linked approach utilizing a subset of precise-distance pulsars restores Earth-pulsar phase coherence. This locks the geometry of the passing wavefront via rapid angular interference oscillations, dramatically refining sky localization.
In the representative injection studied here, the phase-linked search localizes the source to $19.09~{\rm deg}^2$ and $5.46~{\rm deg}^2$ at the 90\% and 50\% credible levels, respectively.

As continuous-wave searches mature into the Square Kilometer Array era~\citep{Shannon_2025}, and as electromagnetic observatories continue to sharpen the astrophysical priors on early-Universe black-hole masses, this phase-coherent, redshift-aware framework will be essential. It transitions PTAs from placing initial bounding limits to directly testing physical models of supermassive black hole seeding, growth, and binary assembly at cosmic dawn.


{\it Acknowledgments.}---This work is supported by the National Key Research and Development Program of China (No. 2023YFC2206704), the Fundamental Research Funds for the Central Universities, and the Supplemental Funds for Major Scientific Research Projects of Beijing Normal University (Zhuhai) under Project ZHPT2025001.
Part of this work is supported by the ARC Centre of Excellence of Gravitational Wave Discovery (CE230100016).
We thank the referees for very helpful comments.
The Parkes radio telescope (Murriyang) is part of the Australia Telescope National Facility which is funded by the Australian Government for operation as a National Facility managed by CSIRO. We acknowledge the Wiradjuri People as the traditional owners of the Observatory site.
This paper includes archived data obtained through the Parkes Pulsar Data archive on the CSIRO Data Access Portal 
(http://data.csiro.au).


{\it Software: } \texttt{enterprise} \citep{enterprise}, 
\texttt{enterprise\_extensions} \citep{enterprise_ext}, 
\texttt{PTMCMCSampler} \citep{PTMCMCSampler_2017,Vousden_2016},
\texttt{libstempo} \citep{libstempo}, 
\texttt{TEMPO2} \citep{Edwards_2006,Hobbs_2006},
\texttt{healpy} \citep{Zonca_2019_healpy}, 
\texttt{HEALPix} \citep{Gorski_2005}.

\nocite{Wang_2021_J0313,Wolf_2026,Yang_2020_J1007,Banados_2018_J1342,Mortlock_2011_J1120,Willott_2003_J1148,Kurk_2007_J1030,Scheel_2009,Berti_2006}
\bibliographystyle{apsrev4-2}
\bibliography{ref}

\newpage

\appendix

\setcounter{table}{0}
\renewcommand{\thetable}{S\arabic{table}}
\setcounter{figure}{0}
\renewcommand{\thefigure}{S\arabic{figure}}





\begin{center}
{\bfseries Supplemental Material for\\
\textit{Nanohertz Gravitational-Wave Constraints on Supermassive Binary Black Holes at Cosmic Dawn}}
\end{center}

\begin{table*}[t]
    \centering
    \caption{\label{tab:quasar_mass_limits} 
    Properties and PTA constraints for the high-redshift quasar sample. The adopted value is used only to define an equal-mass binary reference chirp mass and should not be interpreted as a unique measurement of the black-hole mass. For each target, the electromagnetically reported central black hole mass ($M_{\rm BH}$) is assumed to be the total mass of a putative binary, and converted to an intrinsic equal-mass ($q=1$) chirp mass to enable direct comparison with the strongest PPTA DR3 all-sky upper limit ($\mathcal{M}_c^{\rm UL}$) evaluated across the nanohertz band. Note that while the electromagnetically derived black hole masses carry significant uncertainties, we neglect them here because the current PTA upper limits remain a factor of $\gtrsim 10$ above the maximum allowed reference chirp mass for the majority of the sample.
    }
    \begin{ruledtabular}  
    \begin{tabular}{c c c c c}
    Target name & $z$ & $M_{\rm BH}\,(M_\odot)$ & $\mathcal{M}_c\,(M_\odot)(q=1)$ & Best all-sky $\mathcal{M}_c^{\rm UL}\,(M_\odot)$ \\
    \hline
    \rule{0pt}{2.6ex}
    J0313$-$1806~\cite{Wang_2021_J0313,Wolf_2026} & 7.642 & $1.63 \times 10^{9}$ & $7.09 \times 10^{8}$ & $8.18 \times 10^{9}$ \\
    J1007+2115~\cite{Yang_2020_J1007} & 7.515 & $1.50 \times 10^{9}$ & $6.53 \times 10^{8}$ & $8.21 \times 10^{9}$ \\
    J1342+0928~\cite{Banados_2018_J1342} & 7.540 & $8.00 \times 10^{8}$ & $3.48 \times 10^{8}$ & $8.20 \times 10^{9}$ \\
    J1120+0641~\cite{Mortlock_2011_J1120} & 7.085 & $2.00 \times 10^{9}$ & $8.71 \times 10^{8}$ & $8.30 \times 10^{9}$ \\
    J1148+5251~\cite{Willott_2003_J1148} & 6.410 & $3.00 \times 10^{9}$ & $1.31 \times 10^{9}$ & $8.45 \times 10^{9}$ \\
    J0100+2802~\cite{Wu_2015,Eilers_2023} & 6.327 & $1.20 \times 10^{10}$ & $5.22 \times 10^{9}$ & $8.47 \times 10^{9}$ \\
    J1030+0524~\cite{Kurk_2007_J1030} & 6.306 & $1.40 \times 10^{9}$ & $6.09 \times 10^{8}$ & $8.47 \times 10^{9}$ \\
    J0148+0600~\cite{Yue_2024} & 5.977 & $7.79 \times 10^{9}$ & $3.39 \times 10^{9}$ & $8.55 \times 10^{9}$ \\
    \end{tabular}
    \end{ruledtabular}
\end{table*}

{\it A: PTA sensitivity to the high-redshift quasar population.}---To contextualize the current nanohertz sensitivity frontier, we evaluate our source-frame all-sky limits against a representative sample of spectroscopically confirmed $z \gtrsim 6$ quasars (\Table{tab:quasar_mass_limits}). While the strongest all-sky chirp-mass upper limits across the nanohertz band cluster consistently near $\mathcal{M}_c^{\rm UL} \simeq 8 \times 10^9 M_\odot$, the equal-mass reference masses inferred from electromagnetic observations of these quasars span a much wider range ($\sim 3 \times 10^8$ to $5 \times 10^9 M_\odot$). This comparison explicitly isolates J0100+2802 as the most immediate astrophysical candidate for direct PTA constraints. Its equal-mass reference chirp mass resides just a factor of order unity below the current global all-sky limit, whereas the majority of the high-redshift population remains presently inaccessible. Consequently, current nanohertz observatories are actively probing the extreme upper tail of the early-Universe black hole mass function, but are not yet sensitive to the bulk population.

{\it B: High-frequency cutoffs and coalescence timescales.}---In Fig.~1 of the main text, we explicitly truncate the CGW constraints at the ISCO frequency and denote the subsequent parameter space bounded by the ringdown frequency. 
The low-frequency boundary of the transition region is determined by the observed ISCO frequency,
$f_{\rm ISCO}=(1+z)^{-1}c^3/(6^{3/2}\pi G M_{\rm tot})$, 
where $M_{\rm tot}$ is the source-frame total mass and the factor $1/(1+z)$ accounts for cosmological redshift. Since the boundaries in Fig.~1 of the main text are shown in terms of the source-frame chirp mass, we convert $M_{\rm tot}$ to $\mathcal{M}_c$ using $M_{\rm tot}=2^{6/5}\mathcal{M}_c$ for an equal-mass binary.

For an equal-mass, non-spinning binary, the final merged remnant retains a mass of $M_f \approx 0.95 M_{\rm tot}$ and a dimensionless spin of $\chi_f \approx 0.686$~\citep{Scheel_2009}. Using the Kerr quasinormal-mode fitting formula for the dominant $(l=m=2, n=0)$ mode~\citep{Berti_2006}, the observed ringdown frequency is given by $f_{\rm rd} \approx (1+z)^{-1} (0.53 c^3 / 2\pi G M_f)$.

For a $M_{\rm tot} = 1.2 \times 10^{10} M_\odot$ binary at the redshift of J0100+2802 ($z=6.327$), the observed ISCO and ringdown frequencies are around $50 \text{ nHz}$ and $206 \text{ nHz}$, respectively. While these frequencies fall within the sensitivity band of PTAs, such short-duration signals are effectively undetectable by current CGW analyses. At these extreme masses, the characteristic timescale of the system ($t_M = G M_{\rm tot}/c^3$) is approximately $16$ hours. Consequently, the entire plunge, merger, and ringdown sequence occurs over a timescale of merely $1$ to $1.5$ years in the observer frame. Because this duration is substantially shorter than the multi-decade observing baseline, the signal lacks the long-lived stability assumed by standard CGW analyses. However, since the GW strain generated during the merger phase is expected to be significantly stronger than during the inspiral, our constraints extended into the frequency band bounded by the ISCO and ringdown scales serve as a useful and conservative reference. Ultimately, directly probing this final coalescence phase for high-redshift binaries requires dedicated GW burst searches, representing a promising frontier for future multi-messenger campaigns.



{\it C: Injection study setup and parameter recovery.}---To assess parameter recovery and sky localization capabilities, we construct a 40-year mock PTA dataset utilizing a 32-pulsar array (comprising the 30 PPTA DR3 pulsars and two precise nearby MeerKAT pulsars, J0636$-$3044 and J2222$-$0137). The observing cadence for each pulsar is extrapolated from its actual times of arrival, supplemented with random temporal jitter to avoid exact periodicity. The TOA uncertainties are resampled from the empirical distribution of the measured timing uncertainties for each pulsar.
The pulsar-specific noise realizations are generated using the corresponding white- and red-noise parameters derived from the real dataset. White noise is injected through the EFAC and EQUAD parameters, while the intrinsic red noise is modeled as a power-law process characterized by the pulsar-specific red-noise amplitude and spectral index.

We evaluate parameter recovery using two distinct Bayesian search strategies, employing identical priors across an eight-parameter GW signal model (detailed in \Table{tab:injection_recovery}). Redshift is not included as an independent sampling parameter. Instead, the posterior samples of $D_L$ are converted to redshift using the distance--redshift relation under the adopted cosmology.

In the standard ``phase-decoupled" search, all pulsar distances are treated as effectively unconstrained nuisance parameters. Because distance uncertainties are much larger than the gravitational wavelength, the pulsar-term phases are completely randomized and uncoupled across the array, leaving source localization reliant entirely on the broad Earth-term antenna response.

Conversely, the “phase-linked" search employs an anchor-pulsar strategy \cite{Wen_2026}. 

This strategy is similar to the CGW localization studies of Kato and Takahashi~\citep{Kato_2023,Kato_2026}. The common physical mechanism is that accurate pulsar distances constrain the pulsar-term phase through the Earth-pulsar time delay, thereby restoring Earth-pulsar-term phase coherence and improving continuous-wave source localization. While those studies focused on generic continuous-wave source localization using idealized PTA configurations, here we apply the same mechanism to a redshift-aware high-redshift SMBBH signal at the sky position and redshift of J0100+2802.
We impose sub-wavelength distance priors ($D_{\rm err} < \lambda_{\rm GW}$) on six pulsars (J0030+0451, J0437$-$4715, J1744$-$1134, J1909$-$3744, J0636$-$3044, and J2222$-$0137). 

The adopted distance uncertainties are $1\,{\rm pc}$ for J0030+0451, J1744$-$1134, J1909$-$3744, J0636$-$3044, and J2222$-$0137, and $0.1\,{\rm pc}$ for J0437$-$4715. For the injected signal frequency, $f_{\rm GW}=10^{-8.5}\,{\rm Hz}$, the corresponding GW wavelength is $\lambda_{\rm GW}=c/f_{\rm GW}\simeq3.1\,{\rm pc}$; hence all six anchor pulsars satisfy the sub-wavelength condition. The distance uncertainties adopted for this analysis are comparable to the currently achieved measurement precision for these six pulsars; see Ref. \cite{Wen_2026} for details on present-day astrometric accuracies.

Within this sub-array, the pulsar-term phase is no longer absorbed by random distance-uncertainty degrees of freedom. Instead, it remains phase-connected to the Earth term, successfully restoring cross-pulsar phase coherence and effectively transforming the sub-array into a phase-coherent interferometer.

The quantitative impact of this phase-linking is explicitly detailed in \Table{tab:injection_recovery}, which compares the injected parameters against the recovered posteriors for both searches. While the phase-decoupled search successfully bounds the amplitude-correlated parameters, restoring phase coherence drastically sharpens the parameter estimation across the board. 

\begin{figure*}[htbp]
    \begin{center}
        \includegraphics[width=1.0\linewidth]{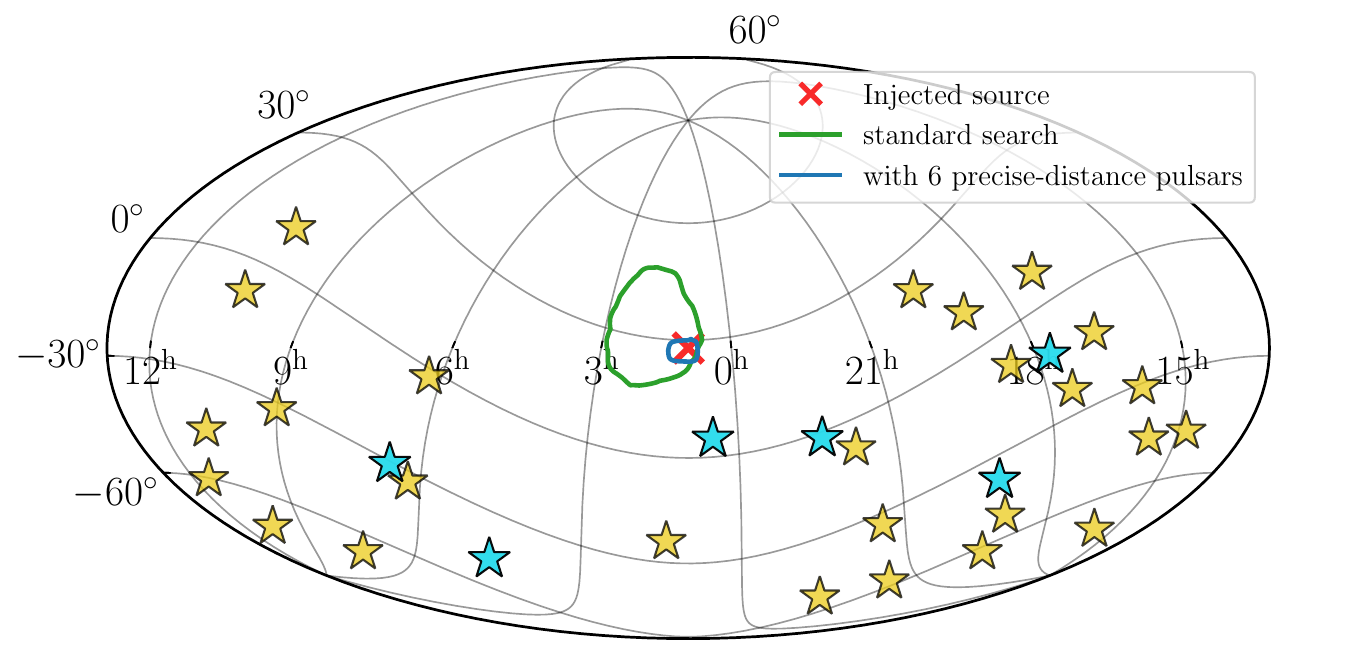}
        \caption{Sky localization of the injected J0100+2802-like SMBBH. Equatorial map showing the 90\% credible regions for the recovered continuous-wave source. The red cross marks the true injected position. The 32 pulsars are denoted by stars, with the specific six precise-distance pulsars utilized in the phase-linked search highlighted in cyan. Solid contours outline the posterior localization areas derived from the standard search (green) and the phase-linked search (blue).} 
        \label{skymap_compare}  
    \end{center}
\end{figure*}

In our analysis, the strongest improvement is in sky localization, while the posteriors of $\mathcal{M}_c$, $f_{\rm GW}$, $D_L$, $\iota$, $\Phi_0$, and $\psi$ are also tightened. This is broadly consistent with Kato and Takahashi~\citep{Kato_2023,Kato_2026}, who found the largest gains for parameters most directly coupled to the pulsar-term phase, particularly the sky position, chirp mass, and GW frequency.
This physical impact is most profoundly visualized in the sky-localization posteriors (\Fig{skymap_compare}). By explicitly locking the relative phase of the passing wavefront, the phase-linked search overcomes the broad spatial uncertainties inherent to the standard analysis, collapsing the 90\% credible sky area from $567.78~{\rm deg}^2$ down to $19.09~{\rm deg}^2$ (and the 50\% credible area from $180.40~{\rm deg}^2$ to $5.46~{\rm deg}^2$). For future multi-messenger campaigns seeking electromagnetic counterparts to high-redshift SMBBHs, this demonstrates that achieving high-precision astrometry for only a small fraction of the PTA pulsars is sufficient to transition CGW searches from broad hemispheric constraints to precision targeted fields.


\begin{table*}[t]
\centering
\caption{
Prior distributions and recovered posteriors for the simulated J0100+2802-like SMBBH. For both the standard phase-decoupled search and the phase-linked (anchor-pulsar) search, we detail the adopted uniform prior ranges, the injected parameter values, and the recovered parameter estimates. 
{
Redshift is not independently sampled. The reported $z$ posterior is derived from the luminosity-distance posterior using the distance--redshift relation under the adopted cosmology~\citep{Planck_2018}.
}
The recovered values are reported as the median of the marginalized posterior, with uncertainties corresponding to the 68\% credible interval. By successfully restoring Earth-pulsar-term phase coherence, the phase-linked search yields substantially tighter constraints across the parameter space, most notably driving the drastic reduction in the sky coordinate (RA and Dec) uncertainties. 
}
\label{tab:injection_recovery}
\renewcommand{\arraystretch}{1.36}
\begin{ruledtabular}
\begin{tabular}{lcccc}
Parameter & Prior & Injected & Standard search & Phase-linked search \\
\hline
RA [deg] 
& $[0,360]$ 
& $15.05425$ 
& $28.218^{+7.629}_{-7.621}$ 
& $16.584^{+1.165}_{-1.178}$ \\

Dec [deg] 
& $\sin({\rm Dec}) \in [-1,1]$ 
& $28.040511$ 
& $31.154^{+7.585}_{-6.259}$ 
& $27.071^{+1.292}_{-1.202}$ \\

$\cos\iota$ 
& $[-1,1]$ 
& $0.800$ 
& $0.602^{+0.254}_{-0.219}$ 
& $0.652^{+0.096}_{-0.135}$ \\

$\log_{10}(\mathcal{M}_c/M_\odot)$ 
& $[8,12]$ 
& $9.718$ 
& $9.613^{+0.162}_{-0.157}$ 
& $9.745^{+0.056}_{-0.036}$ \\

$\log_{10}(f_{\rm GW}/{\rm Hz})$ 
& $[-9,-7]$ 
& $-8.500$ 
& $-8.506^{+0.008}_{-0.008}$ 
& $-8.497^{+0.002}_{-0.002}$ \\

$z$ 
& --
& $6.327$ 
& $2.995^{+1.407}_{-1.042}$ 
& $7.039^{+0.756}_{-0.865}$ \\

$\log_{10}(D_L/{\rm Mpc})$ 
& $[2,6]$ 
& $4.797$ 
& $4.415^{+0.199}_{-0.225}$ 
& $4.851^{+0.051}_{-0.066}$ \\

$\Phi_0$ [rad] 
& $[0,2\pi]$ 
& $0.524$ 
& $3.309^{+1.317}_{-2.578}$ 
& $2.579^{+0.989}_{-0.478}$ \\

$\psi$ [rad] 
& $[0,\pi]$ 
& $0.785$ 
& $1.516^{+0.634}_{-1.196}$ 
& $1.216^{+0.241}_{-0.497}$ \\

\end{tabular}
\end{ruledtabular}
\end{table*}

\end{document}